\documentclass[twocolumn,trackchanges,twocolappendix]{aastex631}

\usepackage{amsmath}
\usepackage{makecell}
\usepackage{graphicx}
\usepackage{amsthm,amsmath,amssymb} 
\usepackage{mathrsfs}
\usepackage{longtable}
\usepackage{subfigure}
\usepackage{rotating}
\usepackage{amssymb}
\usepackage{epsfig}
\usepackage{multirow}
\usepackage{enumerate}
\usepackage{float}
\usepackage[section]{placeins}
\usepackage{url}
\usepackage{soul}
\usepackage{CJK}

\def\apjl{Astrophys.~J.~Lett.}
\def\apjs{Astrophys. J., Suppl. Ser.}

\def\aap{Astron.~Astrophys.}

\begin{document}
\begin{CJK*}{UTF8}{gbsn}

\title{Fast radio bursts generated by coherent curvature radiation from compressed bunches \\ for FRB 20190520B}

\author[0000-0002-6165-0977]{Xiang-han Cui}
%\altaffiliation{Corresponding author}
\affiliation{CAS Key Lab of FAST, National Astronomical Observatories, Chinese Academy of Sciences, Beijing 100101, China}
\affiliation{School of Astronomy and Space Science, University of Chinese Academy of Sciences, Beijing 100049, China}
%\email{cuixianghan@nao.cas.cn}

\author{Zheng-wu Wang}
\affiliation{CAS Key Lab of FAST, National Astronomical Observatories, Chinese Academy of Sciences, Beijing 100101, China}

\author[0000-0003-1908-2520]{Cheng-min Zhang}
%\altaffiliation{Corresponding author}
\affiliation{CAS Key Lab of FAST, National Astronomical Observatories, Chinese Academy of Sciences, Beijing 100101, China}
\affiliation{School of Astronomy and Space Science, University of Chinese Academy of Sciences, Beijing 100049, China}
\affiliation{School of Physical Sciences, University of Chinese Academy of Sciences, Beijing 100049, China}
%\email{zhangcm@bao.ac.cn}

\author[0000-0001-6651-7799]{Chen-hui Niu}
\affiliation{Institute of Astrophysics, Central China Normal University, Wuhan 430079, China}
\affiliation{CAS Key Lab of FAST, National Astronomical Observatories, Chinese Academy of Sciences, Beijing 100101, China}

\author[0000-0003-3010-7661]{Di Li}
%\altaffiliation{Corresponding author}
\affiliation{CAS Key Lab of FAST, National Astronomical Observatories, Chinese Academy of Sciences, Beijing 100101, China}
\affiliation{School of Astronomy and Space Science, University of Chinese Academy of Sciences, Beijing 100049, China}
\affiliation{NAOC-UKZN Computational Astrophysics Centre, University of KwaZulu-Natal, Durban 4000, South Africa}
\affiliation{Research Center for Intelligent Computing Platforms, Zhejiang Laboratory, Hangzhou 311100, China}
%\email{dili@nao.cas.cn}

\author[0000-0003-2357-6259]{Jian-wei Zhang}
\affiliation{CAS Key Lab of FAST, National Astronomical Observatories, Chinese Academy of Sciences, Beijing 100101, China}
%\affiliation{Yunnan University, Kunming, Yunnan, 650500, China}

\author{De-hua Wang}
\affiliation{School of Physics and Electronic Science, Guizhou Normal University, Guiyang 550001, China}

\correspondingauthor{Chengmin Zhang (zhangcm@bao.ac.cn) and Di Li (dili@nao.cas.cn)}

\begin{abstract}
The radiation mechanism of fast radio bursts (FRBs) has been extensively studied but still remains elusive. 
Coherent radiation is identified as a crucial component in the FRB mechanism, with charged bunches also playing a significant role under specific circumstances. 
In the present research, we propose a phenomenological model that draws upon the coherent curvature radiation framework and the magnetized neutron star, taking into account the kinetic energy losses of outflow particles due to inverse Compton scattering (ICS) induced by soft photons within the magnetosphere.
By integrating the ICS deceleration mechanism for particles, we hypothesize a potential compression effect on the particle number density within a magnetic tube/family, which could facilitate achieving the necessary size for coherent radiation in the radial direction.
This mechanism might potentially enable the dynamic formation of bunches capable of emitting coherent curvature radiation along the curved magnetic field.
Moreover, we examine the formation of bunches from an energy perspective.
Our discussion suggests that within the given parameter space the formation of bunches is feasible.
Finally, we apply this model to FRB 20190520B, one of the most active repeating FRBs discovered and monitored by FAST.
Several observed phenomena are explained, including basic characteristics, frequency downward drifting, and bright spots within certain dynamic spectral ranges.

%In particular, this model may also explain the position of bright spots in the dynamic spectral diagram.
%which we suggest that this may be related to the shape of the particle outflow region.
%Additionally, we consider the kinetic energy loss of outflow particles due to the inverse Compton scattering (ICS) caused by soft photons in the magnetosphere.
%the compression effect on the particle number density is realized in a magnetic tube/family, which will achieve the required size for coherent radiation in the radial direction.
%However, the formation of bunches is not clear enough, even though the hypothesis of two-stream instability has been proposed.
%because we lack accurate simulations of the magnetosphere of neutron stars under extreme relativistic conditions.

\end{abstract}


\section{Introduction} \label{sec:intro}

The systematic researches of distant fast radio bursts (FRBs), a class of bright millisecond radio transients, have started in 2007 \citep{Lorimer07}.
Throughout 16 years of accelerated development, numerous significant observational milestones have been accomplished \citep{Spitler16, Chatterjee17, Bochenek20,CHIME20a}, which are comprehensively reviewed in the literature \citep{Cordes19,Petroff22}. 
To date, over 700 FRBs \citep{CHIME21a} and 5,000 bursts\footnote{https://blinkverse.alkaidos.cn} have been recorded, and several expansive datasets corresponding to individual sources have been published \citep{Li21b, Niu22, Xu22}.
In addition, the abundant data also provide a foundation for statistical studies of FRBs \citep{Feng22}, such as FRB luminosity function \citep{Luo18, Hashimoto22}, FRB population analysis \citep{Cui21, Pleunis21}, FRB as cosmological probes \citep{Macquart20,James22b}, etc.

Despite rigorous exploration, the precise origin and radiation mechanism of FRBs remain enigmatic. 
Regarding their origin, the magnetar model \citep{Popov10} garners support from observations of the Galactic SGR 1935+2154 \citep{Bochenek20,CHIME20a}. 
However, no such bright bursts have been detected in subsequent observations \citep{Lin20}, and comparable radio bursts have not been exhibited by other magnetars.
As for their radiation mechanism, coherent radiation has emerged as the most feasible explanation for FRBs, given their exceedingly high brightness temperature \citep{Lu18, Lyutikov21}, yet the exact physical processes are continue to be a topic of considerable debate.

Broadly speaking, the mechanism of FRBs can be divided into two categories:  inside the magnetosphere (pulsar-like) and outside the magnetosphere (GRB-like), for detailed information, can see the model review \citep{Lyubarsky21, Xiao21, Zhang22}.
Among these, the model proposing coherent curvature radiation by bunches \citep{Katz14, Cordes16, Kumar17, Ghisellini18, Yang18, Wang20, Cooper21} represents one of the foremost contenders for explaining FRBs.
These models posit that FRBs stem from the curvature radiation generated by relativistic bunches traversing along the curved magnetic field lines.
Observational constraints stipulate that the size of these bunches ($l_b$) should be less than half of the wavelength (e.g., $\lambda /2 \sim 10\,cm$ for 1GHz) to ensure that the radiation signals from various charged particles within a single bunch reach the observer in approximately the same phase \citep{Wang20}.
This phenomenon is thus termed coherent curvature radiation.

Nonetheless, these models confront two primary obstacles. 
The initial challenge pertains to the formation of relativistic bunches \citep{Melrose17}, and the secondary challenge relates to the influence of plasma effects \citep{Lyubarsky21, Beloborodov21}. 
In addressing the latter issue, \cite{Qu22} demonstrated that bright FRBs could escape the magnetosphere across a vast parameter space and propagate outward.
But, the first issue still lacks a satisfactory resolution.
Although the hypothesis of a two-stream instability has been posited \citep{Usov87, Melikidze00}, simulation studies to corroborate this conjecture are absent under extreme relativistic conditions \citep{Samuelsson10, Tokluoglu18}.
More specifically, the formation problem is attributed to the possibility that Coulomb forces, electrostatic potential, or other interactions could cause dispersion among charged particles within the bunch, thereby inhibiting their congregation \citep{Zhang22}.

In this paper, we aim to analyze the formation of dynamic bunches through the inverse Compton scattering (ICS) mechanism, which precedes curvature radiation and leads to additional energy loss that cause compression of the outflow stream.
Building on this model, we intend to further investigate the observational features of FRB 20190520B.
This FRB was detected by the Five-hundred-meter Aperture Spherical radio Telescope (FAST) \citep{Li18, Niu22} and is distinct not only due to its association with a persistent radio source (PRS) (the other is FRB 20121102A), but also owing to its sustained activity since its discovery (Niu et al. 2023 in prep.). 
This allows for ongoing monitoring and the discovery of new features, and thus we have selected FRB 20190520B as a case study for our modeling and analysis.
In Section \ref{2}, we establish the model from the perspective of the bunch compression process. 
In Section \ref{3}, we present simulations based on our model, applying this model to FRB 20190520B, and providing explanations for observed phenomena. 
Subsequently, in Section \ref{4}, we discuss other related implications and concerns. 
Finally, we summarize our conclusions in Section \ref{5}.

%Therefore, one key point of our model is how these bunches were formed, which will be explained later.
%With the completion and commissioning of several powerful telescopes, such as Canadian Hydrogen Intensity Mapping Experiment (CHIME) \citep{CHIME18}, Australian Square Kilometre Array Pathfinder (ASKAP) \citep{Macquart10}, and  Five-hundred-meter Aperture Spherical radio Telescope (FAST) \citep{Li18}, the observational information of FRBs significantly increase.

\section{Compressed bunch Model} \label{2}

In this section, we will construct our model from a physical perspective. 
Initially, a general overview of the model will be provided, followed by the introduction of three model components. 
Subsequently, we present the derivation of the model and parameters, underpinned by plausible assumptions related to neutron stars (NSs).

\subsection{General illustration}\label{2.1}

\begin{figure*}[ht]
\plotone{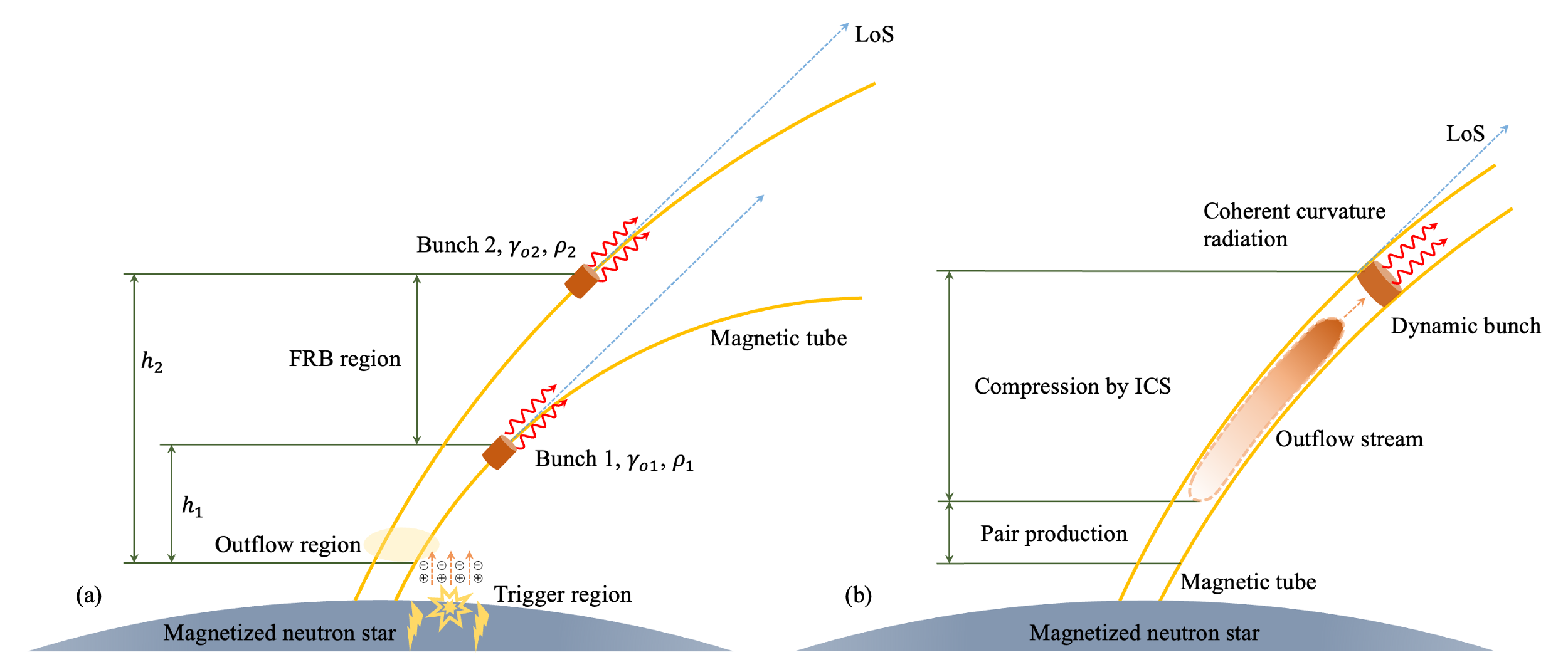}
\caption{Schematic diagram of one burst based on the compressed bunches model and coherent curvature radiation from a magnetized neutron star (NS).
In sub-figure (a), the brown cylinders represent various bunches with different heights ($h_1$ \& $h_2$), Lorentz factors ($\gamma_{o1}$ \& $\gamma_{o2}$), and curvature radii ($\rho_1$ \& $\rho_2$).
The red wavy lines signify the coherent curvature radiation emitting from the bunches.
The yellow curves depict the magnetic tubes, and the blue dashed lines represent the line of sight (LoS).
Moreover, the regions of trigger, outflow, and FRB emissions are also displayed.
In sub-figure (b), the striped area circled by the orange dashed line represents the outflow stream that undergoes compression via inverse Compton scattering (ICS), leading to the formation of the dynamic bunch.
The dark and light orange part represent the upstream and downstream of the outflow particles, respectively.}
The regions of pair production and ICS are also marked.
\label{model}
\end{figure*}

In our model, we consider the whole process of FRB emission, drawing upon the framework of coherent curvature radiation by bunches and the fundamental physics of NS \citep{Lorimer12, Lyne12}.
Generally speaking, this kind of model should depict at least three components, as shown in Figure \ref{model}, which encompass the original energy source (trigger region), the formation of bunches (outflow region), and the radiation ensuing from the movement of bunches or interaction with the surrounding medium (FRB region).

Firstly, the trigger mechanisms for FRBs are generally associated with several instances of intense energy release, such as crust cracking of the NS surface \citep{Thompson01, Beloborodov07}, sudden magnetic reconnection \citep{Lyubarsky20}, or the discharge of magnetic energy following glitches in the NS \citep{Wang21}.
Thus, all these mechanisms can be considered as being magnetically powered.
These events, which we regard as feasible in the context of astrophysics, could result in an immense outflow of relativistic charged particles.
Theoretically, such events need to occur outside of the outflow magnetic field lines, such as beyond the open magnetic field region \citep{Lu20}.
If this is not the case, the violent eruptions could disrupt the ordered magnetic field. 
Therefore, only a minor quantity of charges will enter the open field region and stream out along the magnetic field following pair production, providing the seed charges for FRBs.

Secondly, the formation of bunches poses a key question in this type of model \citep{Kumar17}.
In this context, we present a potential scenario for the formation of bunches.
In contrast to other bunched coherent curvature radiation mechanisms, we postulate that other radiation processes precede FRB emission \citep{Longair10}, including pair production, bremsstrahlung, and inverse Compton scattering (ICS).
These processes are situated closer to the NS surface (in the outflow region) than the curvature radiation of FRB (FRB region), as depicted in Figure \ref{model}.
Given the considerations of mean free path (acting distance) \citep{Timokhin19} and scattering cross-section, we argue that ICS is the predominant process (refer to Appendix \ref{B} for further details). 
Therefore, before reaching the point of FRB emission, the outgoing relativistic charged particles undergo energy loss due to ICS.

Assuming an environment saturated with soft photons originating from the NS surface and pervading the inner magnetosphere, the energy loss rate of a relativistic particle becomes significant. 
Consequently, its kinetic energy experiences a considerable reduction even over a short period of ICS action.
Simultaneously, for the upstream particles (the preceding portion of the outflow stream, represented by the darker segment in sub-figure (b) of Figure \ref{model}), the duration of ICS is extended, resulting in a more pronounced energy loss compared to the downstream particles (the subsequent portion of the outflow stream, represented by the lighter segment).
Macroscopically, the stream undergoes compression as it moves, tending towards unification for six dimensions (three spatial and three momentum dimensions). 
This process may facilitates the dynamic formation of bunches.

Finally, the compressed relativistic bunches traverse along the curved magnetic field, emitting coherent curvature radiation in the tangent direction with various heights.
The frequencies of this radiation are correlated with the distinct radii of curvature and the Lorentz factor ($\gamma$) unique to each bunch.
Among these, the number density of the compressed bunch may be exceedingly high, thereby a magnetic tube/family should be considered, which is related to the state of NS \citep{Link03, Lander13}.
It is crucial to emphasize that the bunches are dynamically compressed during their movement and radiate at certain instants, without the need for long-term stability.
The previously mentioned content provides a phenomenological description of the compressed bunch model, and we will now proceed to outline the physical derivation of this model.

%Furthermore, considering that the size of one compressed bunch is $\sim 10\, cm$, the corresponding emission time is $\sim 1 \,ns$.
%Thus, the difficulty in maintaining bunches can be avoided, because we only need the bunches to be dynamically compressed during their movement and emit radiation at a certain moment, while it does not need the compressed bunches to maintain for a long time.

\subsection{Physical derivation} \label{2.2}
Let us start with the first component: from the energy budget to the estimation of total particles.
Assuming we interpret the central engine as a magnetized compact object, such as a magnetar \citep{Duncan92}.
Then, we can compute the approximate total magnetic energy within a specific volume as $E_{tot} \approx B_s^2V_{m}/(8\pi)$, where $B_s \sim 10^{15}\, G$ denotes the surface magnetic field, and $V_{m} \sim 3\times10^{13}\,cm^3$ signifies a volume that can completely release its magnetic energy (refer to Appendix \ref{A} for parameter estimations).
%If we believe that the FRBs are magnetic-powered or other explosions induce the release of magnetic energy, 
The $E_{tot}$ should constitute the upper limit of a single FRB burst.
The released energy surges out of the trigger region as photons or electrons with an extraordinarily high Lorentz factor $\gamma_{6}\sim 10^6$ \citep{Ruderman75, Medin10}.
Then the total number of particles is,
\begin{equation}
\begin{split}
    N_{tot} = \eta B_s^2V_{m}/(8\pi \gamma m_e c^2)\sim 10^{42} B_{s,15}^2V_{m,13}\gamma_6 ^{-1},
    \label{ntot}
\end{split}
\end{equation}
in which $\eta$ is efficiency factor of energy conversion, $m_e$ is the mass of an electron, and $c$ is the speed of light.
If the conversion rate of this process is close to nuclear energy release, then the $\eta \sim 10^{-3}$.

However, considering that the trigger region is outside the outflow region, and not all outflow particles will be coherent in the FRB region, some constraint factors need to be further added.
For example, $\zeta$ is the particle number rate from the trigger to the outflow region $\sim 10^{-3}$; also, $\xi$ means the volume factor, which will be given later.
Meanwhile, considering that the tube is not full of the whole outflow region, so another factor is $\kappa \sim 10^{-8}$ (details see Appendix \ref{A}).
%that only the magnetic tubes can confine so many particles, and only particles in the tubes can be coherent
Therefore, the number of seed particles that can coherent is
\begin{equation}
\begin{split}
    N_{tc} &=  \kappa \eta \zeta \xi B_s^2V_{m}/(8\pi \gamma m_e c^2)\\
    &\sim 1.5 \times 10^{28} \xi B_{s,15}^2V_{m,13}\gamma_6 ^{-1},
    \label{ntoc}
\end{split}
%\kappa \zeta \xi N_{tot}
\end{equation}
where $\xi$ is given in Eq.\ref{xi}.
When these ultra-relativistic particles flow out, their high Lorentz factor cannot be maintained because of the pair production process \citep{Ruderman75}.
The cascade is not the key question of this article, so we will only describe it phenomenologically.
Generally, the change of the $\gamma$ is related to the cascade number, and the pair production becomes the main factor in the decrease of $\gamma$, such as in the condition of $\gamma_6$.
However, as the $\gamma$ decreases to $\gamma_3 \sim 10^3$ \citep{Longair10}, the effect of ICS becomes significant (see Appendix \ref{B}).
Thus, the cascade number is estimated as \citep{Timokhin19} 
\begin{equation}
    N_{cas} = log_2(\gamma_6 / \gamma_3)\approx 10,
    \label{ncas}
\end{equation}
and the particle number after the pair production is $N_{e} = 2^{N_{cas}} N_{tc}$.
%\begin{equation}
%   N_{te} = 2^{N_{cas}} N_{tc} \sim 2^{10} \kappa \eta \zeta \xi B_s^2V_{m}/(8\pi \gamma_6 m_e c^2).
%  \label{ne}
%\end{equation}
Here, we have roughly completed the first part of the model that the estimation of particle number for the energy budget.

Next, the second part focuses on the energy loss and compressed mechanism to dynamically form the relativistic bunches.
As we considered in Section \ref{2.1}, particles have energy loss mechanisms during their outflow process, and the energy variation can be expressed through changes in the $\gamma$.
If we believe that the upstream particles lose energy for a longer time than the downstream particles, then the upstream part will experience a greater energy loss and have a smaller $\gamma$.
In other words, there will be a speed difference between the upstream and downstream, and a distance change will occur after some time. 
Specifically, this manifests as a reduction in the distance difference between them, and the downstream will move towards the upstream, resulting in a compression effect, as illistrated in Figure \ref{model}.
%If we consider the compression is due to the energy loss of particles, 
Then the compressed length is written as
\begin{equation}
    \delta l \sim (v_i - v_o) \delta t = (v_i - v_o)(\gamma_i - \gamma_o) m_e c^2/\dot E_{loss},
    \label{deltal}
\end{equation}
where $\delta t$ is the compression time, $v$ is the velocity of electrons, subscripts $i$ and $o$ represent the parameters at the beginning (in) and end (out) of the compression, respectively, and $\dot E_{loss}$ is the power of energy loss mechanism.
\begin{figure}[ht]
\plotone{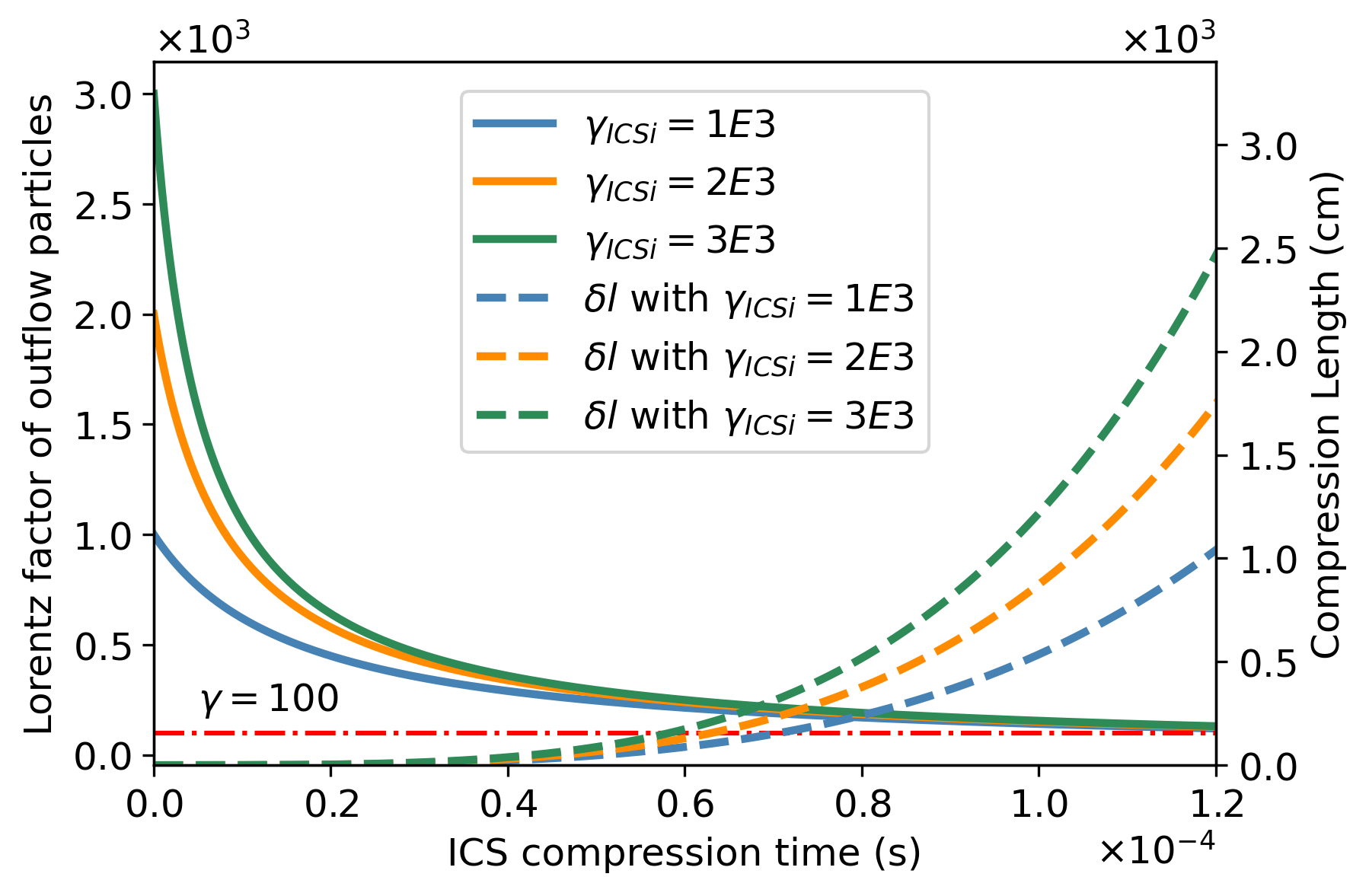}
\caption{Evolution path of the Lorentz factor ($\gamma$) and compression length of the outflow particles with time ($t$) due to the inverse Compton scattering (ICS) compression effect. 
The blue, orange, and green solid lines represent the initial $\gamma_{ICSi}$ with different values at the beginning of ICS energy loss (see left ordinate).
The blue, orange, and green dashed lines represent the compression length ($\delta l$) with different values of $\gamma_{ICSi}$ (see right ordinate).
The red dash-dot line means $\gamma=100$.
\label{tgl}}
\end{figure}
As analyzed above, the ICS is dominated during this process, which power for a single electron can be calculated as \citep{Longair10} 
\begin{equation}
    P_{eic} = \frac{4}{3} \sigma_T c \gamma_o^2 \beta^2 U_{ph},
    \label{peic}
\end{equation}
where $\beta = v/c \approx 1$ under the relativistic condition, and $U_{ph} = \sigma_{sb} T^4/c$ is the internal energy of photons with a constant temperature distribution $T \sim 10^{6} \, K$ \citep{Goldreich69} (surface temperature of a NS) and Stefan-Boltzmann constant $\sigma_{sb}$.
If we consider that $\gamma$ varies with time, the Eq. \ref{peic} can be written as follows,
\begin{equation}
    P_{eic} = \dot E_{loss} = m_e c^2 \dot \gamma_o,
    \label{peicg}
\end{equation}
and by solving the equation, $\gamma_o(t)$ can be expressed as
\begin{equation}
    \gamma_o(t) = \frac{m_e c}{4/3 U_{ph} \sigma_T (t+t_0)}.
    \label{gammat}
\end{equation}
When we choose $\gamma_{o,3} = 10^{3}$ as the $\gamma$ at the ICS initial point of $t=0$, the constant $t_0 = 1.6 \times 10^{-5}\,s$.
Subsequently, the relation between $\gamma$ and ICS compression time can be obtained, as shown in Figure \ref{tgl}. 
From the figure, we can observe that regardless of the initial $\gamma$, when the compression time approaches around $10^{-4}\,s$, their values are approximately 100.
Meanwhile, since it is generally believed that $\gamma$ of coherent radiation in the FRB region is $\sim 10^1-10^2$ \citep{Kumar17, Wang20},  we take its value as $\gamma_{o,2} \sim 100$.
Furthermore, from Figure \ref{tgl}, we also notice that at a compression time of $10^{-4}\,s$, the compression length is $\sim 5\times10^{2}\,cm$. 
Additionally, it can be found that for different initial $\gamma_{ICSi}$, the compression length remains $\sim 10^{3}\,cm$.
Based on these results, we can incorporate the values of $\gamma_{o,2} \sim 100$, $\delta t \sim 10^{-4}\,s$, and $\delta l \sim 5\times10^{2}\,cm$ into the calculations.
%If $\gamma_i \sim \gamma_3$ and $\gamma_o$ are taken, then let $\dot E_{loss} = P_{eic\gamma_i}$ for one electron, and we can estimate the length of compression by Eq. \ref{deltal}, which is $\delta l \sim 2.3\times10^{2}\,cm$.
%Another method to estimate $\delta l$ is using $\delta t$, which we can assume it as the electron lifetime of ICS, 
Because the $\delta l \gg l_{b}$, where $l_{b} \sim 10\,cm$ is the bunch size, $\delta l$ can be regarded as the length of the particle outflow before compression that $l_c \sim \delta l$.
Here, the physical interpretation of $l_c$ is that it represents the portion of the total particle outflow that can be compressed into the size of the bunch.
Then we can use $l_c$ to evaluate the volume factor $\xi$,
\begin{equation}
    \xi = (S_{tube} l_c)/(S_{tube} l_{str}) = l_c/(c t_{tri}),
    \label{xi}
\end{equation}
where $l_{str}$ is the length of the total particle stream, and $t_{tri}$ is the duration of the trigger burst.
If the duration of the trigger is $t_{tri} \sim 10\,s$ like neutron star glitch \citep{Ashton19}, then $\xi \sim 7.7 \times 10^{-10}$.
Through this process, a particle stream with $\sim 5\times10^{2}\,cm$ can potentially be compressed to $10\,cm$ in length.
Now, we can say that the bunch is dynamically formed, and the number of particles of each bunch can be estimated as
\begin{equation}
\begin{split}
    N_{e} \sim 2.5 \times 10^{22} B_{s,15}^2V_{m,13}\gamma_6 ^{-1}.
    %&\sim 10^{-19} B_s^2V_{m}/(8\pi \gamma_6 m_e c^2)\\
\end{split}
\label{ne}
\end{equation}
Correspondingly, its number density is 
\begin{equation}
    n_{b} = N_{e}/(\gamma_o l_b S_{tube}) \sim 2.5\times10^{17} N_{e,22} \gamma_{o,2}^{-1}\,cm^{-3}.
    \label{nb}
\end{equation}
%where $S_{tube}$ is the area of the magnetic tube at the FRB region, if we take $l_b \sim 10\,cm$ and $S_{tube2} = 25 \pi\,cm^2$.

In the third part, we will calculate the luminosity and energy for coherent curvature radiation.
For a single electron, the power of curvature radiation \citep{Longair10} is $P_{ecr} = 2e^2 \gamma_o^4 c/(3\rho^2)$.
So, for one bunch, the power of coherent curvature radiation is
\begin{equation}
    P_{bcr} = N_e^2 P_{ecr} = \frac{2}{3}N_e^2 e^2 \gamma_o^4 c \rho_7^{-2},
    \label{pbcr}
\end{equation}
in which $\rho_7 \sim 10^7\,cm$ is the radius of curvature.
If multiple bunches ($N_b\sim 10^4$) are involved in radiation, then they are incoherent with each other, which power can be written as
\begin{equation}
    P_{cr} = \frac{2}{3}N_b N_e^2 e^2 \gamma_o^4 c \rho_7^{-2}.
    \label{pcr}
\end{equation}
Because of the transformation of the coordinates \citep{Kumar17}, the isotropic luminosity is  $L_{iso} = \gamma_o^4 P_{cr}$ that
\begin{equation}
    L_{iso} \sim 3.0\times 10^{42}N_{b,4} N_{e,22}^2 \gamma_{o,2}^8 \rho_{7}^{-2}\,erg\,s^{-1}.
    \label{liso}
\end{equation}

Regarding the duration of FRBs, as our model is also based on the curvature radiation, some models have already analyzed it thoroughly \citep{Wang20}. 
Therefore, we only provide a brief analysis here that the duration is related to the height difference ($\delta h$) of multiple bunches as $t_{FRB} \sim \delta h/ c\sim 1\,ms$, and $\delta h$ is ten times the radius of a NS.
\begin{equation}
    \delta h = h_2 - h_1 \leq \rho_2 - \rho_1 \sim 10^7\,cm,
    \label{detlah}
\end{equation}
where $h_1$ ($h_1$) and $\rho_1$ ($\rho_2$) represent the bunch at position 1 (2), respectively.

The characteristic frequency \citep{Longair10} of the curvature radiation is related with $\gamma$ and $\rho$ that
\begin{equation}
    \nu_c = 3 c \gamma_o^3/(4 \pi \rho) \approx 0.7 \gamma_{o,2}^3 \rho_{7}^{-1}\,GHz.
    \label{nu}
\end{equation}
Due to multiple bunches, various $\gamma_o$ exist, which will move along with magnetic tubes that have different $\rho$.
Hence, the maximum radiation frequency bandwidth for one burst is 
\begin{equation}
    \Delta \nu_{max} \sim \nu_1 - \nu_2 = \frac{3c}{4\pi} \frac{\rho_2 \gamma_{o1}^3 - \rho_1 \gamma_{o2}^3}{\rho_1 \rho_2},
\end{equation}
$\gamma_{o1}$ and $\gamma_{o2}$ are Lorentz factors of bunches at the lowest and highest positions within the FRB region (see Figure \ref{model}), respectively.
It should be noted that the $\Delta \nu_{max}$ here is not the observed bandwidth but rather the theoretical bandwidth of a single burst. 
In reality, only a portion of the bandwidth may be observed due to the signal-to-noise ratio.
Additionally, the time of bunches emission is also sequential, so there may be frequency drift, which will discuss later.
Above all, the structure of our model has been completed, and we will apply it to the FRB 20190520B in the next Section.
%which is related to the next Section with FRB 20190520B.

\section{Results and Application} \label{3}

In this section, we will utilize the model derived in the preceding section to illustrate its performance under various parameters and provide associated inferences.  
Initially, considering our model as a dynamic mechanism incorporating parameters such as $\gamma$ and $\rho$, we will present a series of inferences based on the model. 
Then, we will conduct basic simulations to enrich our understanding. 
Subsequently, these results will be compared with the actual observations of FRB 20190520B, with the aim of elucidating several phenomena, including but not limited to frequency drifting and morphology.

\subsection{Model inferences}\label{3.1}
Due to the multiple bunches involved in FRB radiation, each bunch contributes to the total radiation.
However, in Eq.\ref{liso}, we assume that their contributions are equal (as in most other models).
This assumption may not reflect physical reality accurately as different bunches may originate from varying heights and $\gamma$.
The existence of these different parameters is roughly what causes the burst duration to broaden to approximately $\sim 1\,ms$.
As such, it becomes important to examine the properties of individual bunches and their contributions to the FRB emission.
\begin{figure}[htbp]
\plotone{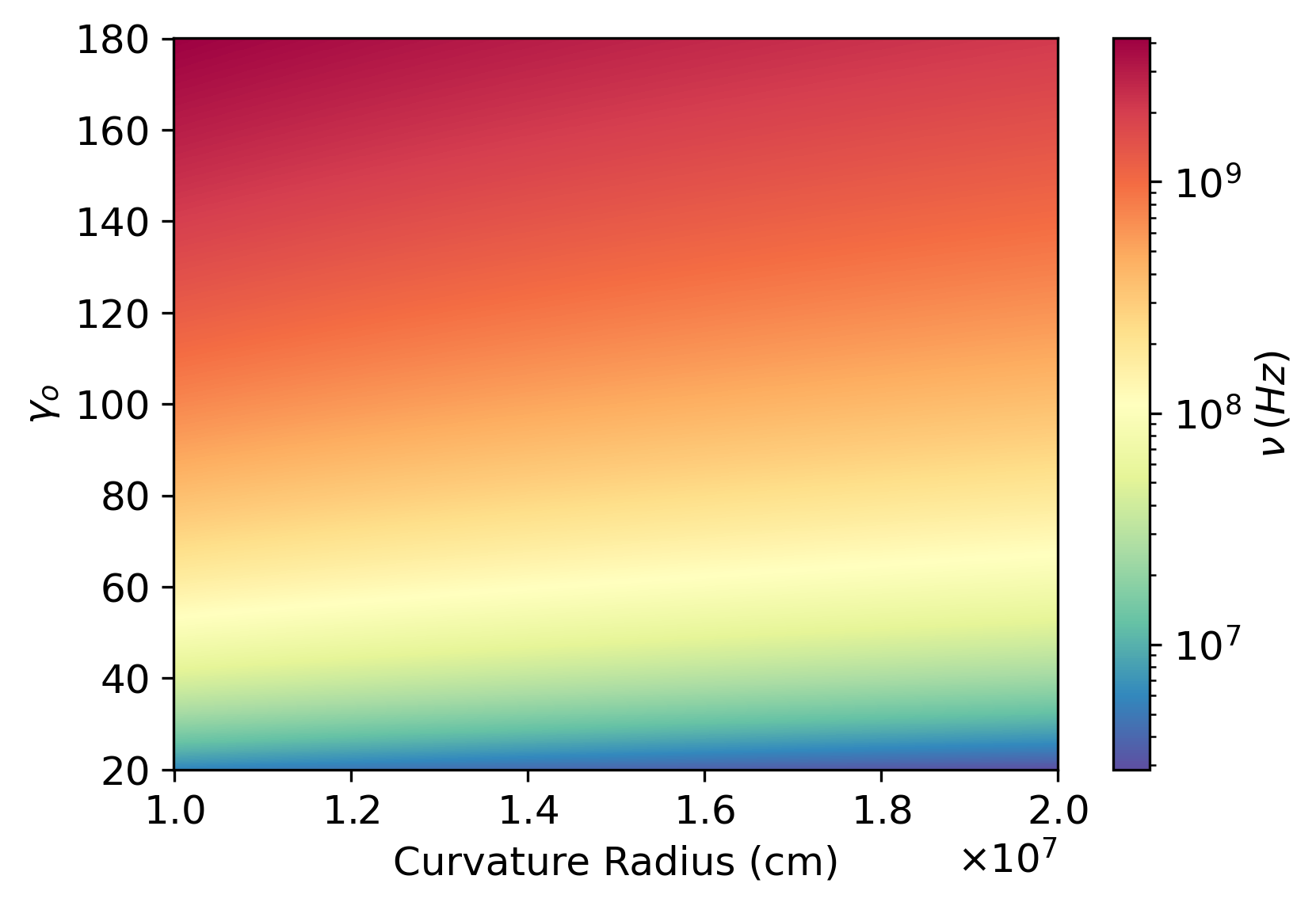}
\caption{The characteristic frequency distribution diagram of a single bunch under different Lorentz factors ($\gamma_o$) and curvature radii ($\rho$).}
\label{rgv}
\end{figure}

\begin{figure}[htbp]
\plotone{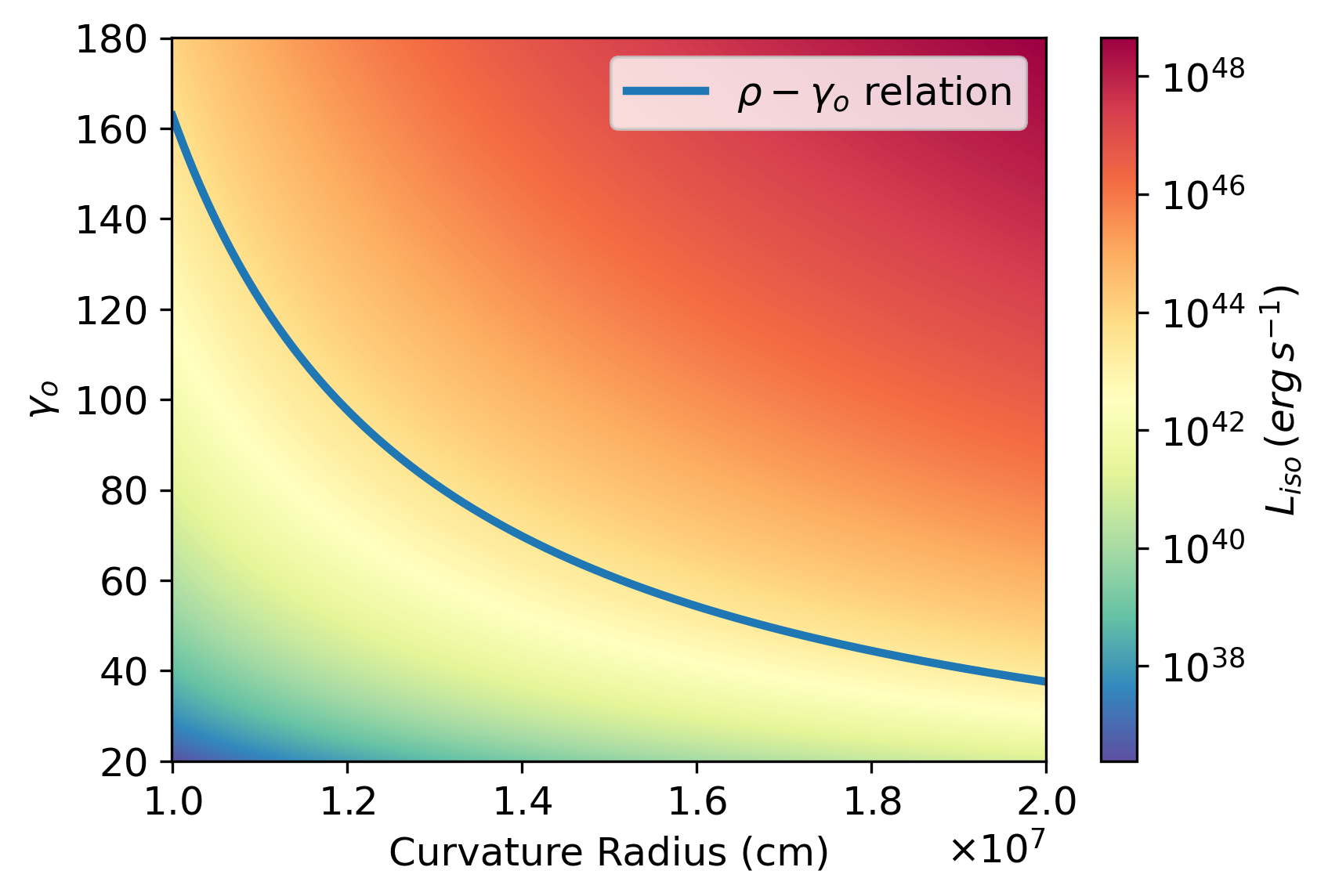}
\caption{The luminosity distribution diagram of a burst (contains multiple bunches) with different Lorentz factors ($\gamma_o$) and curvature radii ($\rho$).
The blue curve represents the relation between $\rho$ and $\gamma_o$ when considering that the value of $\gamma_o$ will decrease as the $\rho$ increases.}
\label{rgl}
\end{figure}

Considering a single bunch, its characteristic frequency ($\nu_c$) and luminosity ($L_{isob}$) can be computed using $Ne$, $\gamma_o$, and $\rho$ according to Eq.\ref{pbcr} and Eq.\ref{nu}.
It's plausible that the higher the launching position of a bunch (i.e., larger $\rho$), the farther it travels, leading to longer energy loss and consequently, a smaller $\gamma_o$ value.
Given these parameters, the variation of $\nu_c$ is graphically represented in Figure \ref{rgv}.
As depicted in Figure \ref{rgv}, as $\gamma_o$ increases and $\rho$ decreases, $\nu_c$ correspondingly increases.
Concurrently, Figure \ref{rgl} illustrates the distribution of $L_{iso}$ across varying values of $\gamma_o$ and $\rho$.
But, as $\gamma_o$ and $\rho$ increase, $L_{iso}$ also increases, which may appear counter-intuitive.
This anomaly arises due to the fact that both $\rho$ and $\gamma_o$ are provided in the simulation code, and the region above the $\rho-\gamma_o$ curve is physically implausible, because when the launching position is higher, the $\gamma_o$ should be smaller.
Hence, only the area beneath the curve can accurately reflect the actual physical conditions.
Having elucidated the properties of individual bunches, we can now proceed to scrutinize the characteristics of a single burst that encompasses multiple bunches.

\subsection{A concise simulation}\label{3.2}

\begin{figure}[!t]
\plotone{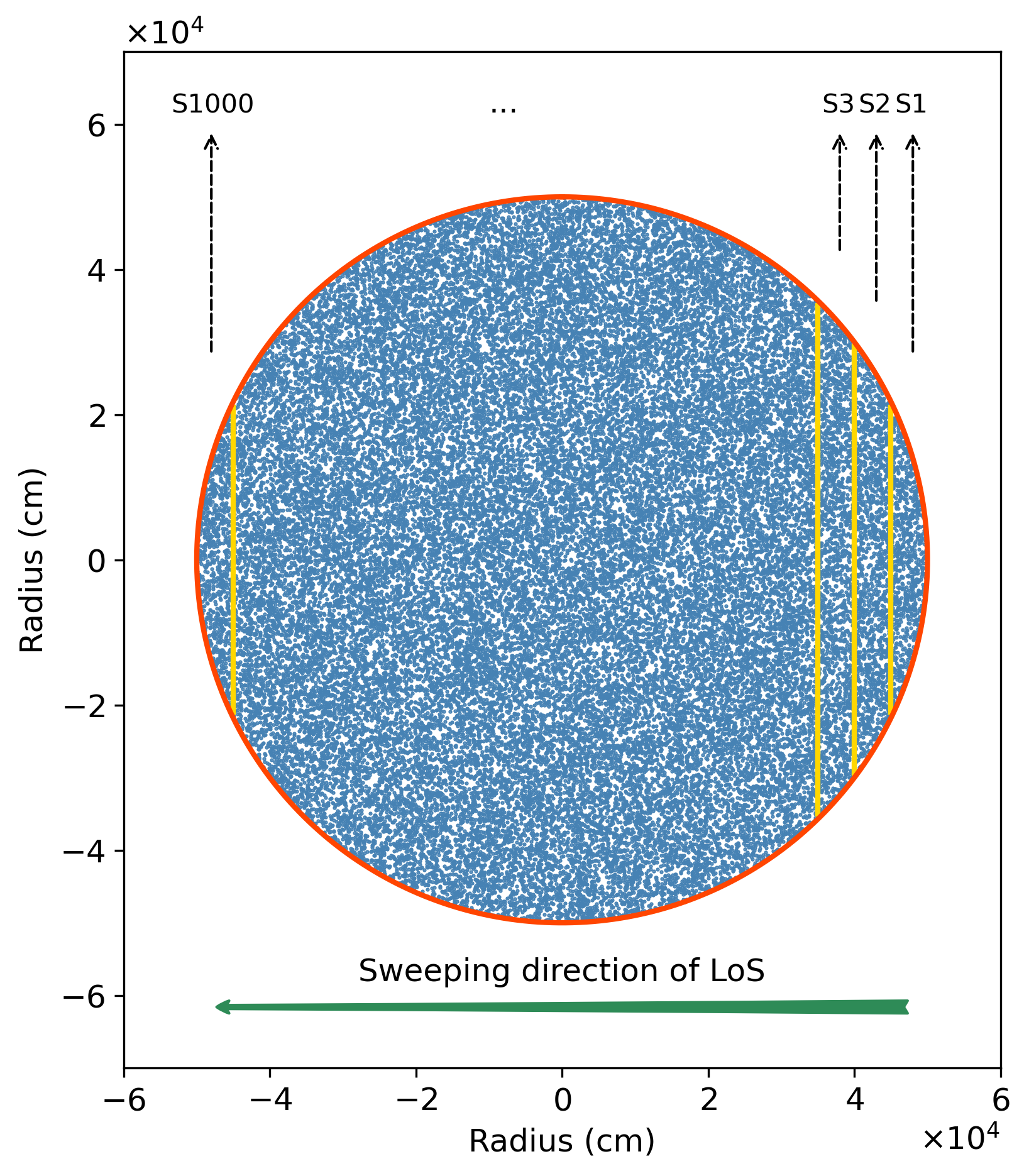}
\caption{Schematic diagram of the slicing method.
Each blue point represents a magnetic field line/tube, which can also be understood as an outflowing bunch.
The area between two yellow dividing lines represents a slice, and there are a total of 1000 slices in the figure, labeled as S1, S2, S3, ..., and S1000.
The dark green arrow represents the direction of the line of sight (LoS), which is the order of the slices.
\label{cycle}}
\end{figure}

\begin{figure}[htbp]
\plotone{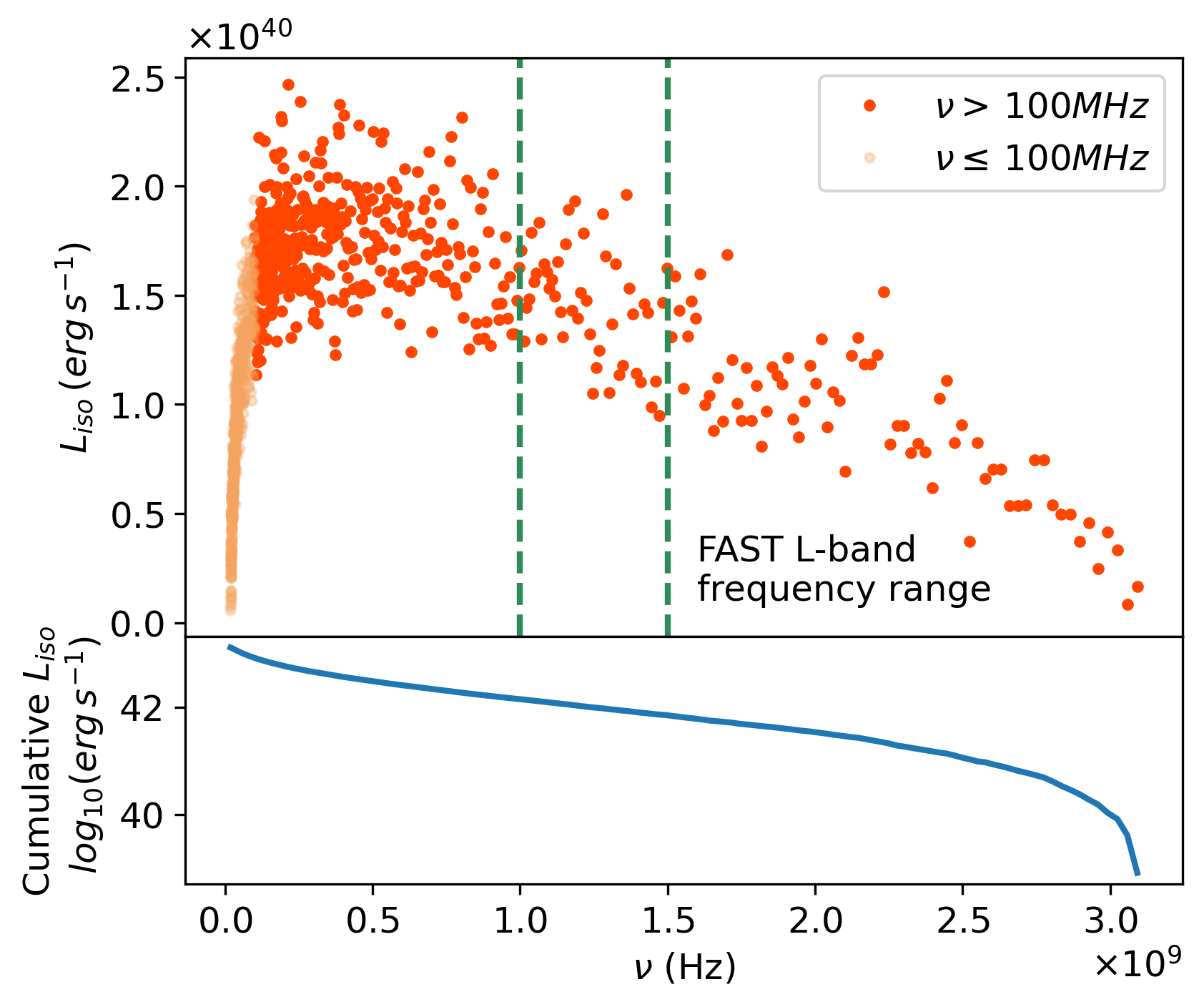}
\caption{The frequency-luminosity diagram for a single burst. 
Upper panel: dots represent different slices (contains multiple bunches), where red-orange points indicate frequencies higher than 100 MHz and light-orange group meaning those below 100 MHz.
Bottom panel: The solid blue line represents the cumulative luminosity.
Note that the vertical axis is presented in a logarithmic scale for ease of reading.
\label{v-L}}
\end{figure}

In this subsection, we will introduce the concept and method of how multiple bunches in different slices are simulated.
If we assume that the outflow region is circular, and the magnetic field lines/tubes are uniformly distributed within it, then we can regard that the curvature radius of adjacent lines/tubes is similar.
First, we randomly distribute points within a circle of radius $R_f$ (outflow region radius) using the Monte Carlo method, with the total number of dots equaling the number of bunches like $N_b\sim 10^4$ and each point represent a bunch, as shown in Figure \ref{cycle}.
Then, we slice the circle uniformly into bins along the direction of the line of sight, for example, from right to left, and count the number of points (bunches) in each slice.
After uniformly dividing the emission height, we can calculate several properties of the bunches, such as compression length (Eq.\ref{deltal}), coherent particle number (Eq.\ref{ne}), isotropic luminosity (Eq.\ref{liso}), and characteristic frequency (Eq.\ref{nu}).
When we assume that the Lorentz factor and curvature radiation luminosity of the bunches in one slice is the same in the FRB region, we use the sum of the luminosity of all bunches to represent the luminosity of a slice (one dot in Figure \ref{v-L}).
Finally, these luminosities and characteristic frequencies are plotted in Figure \ref{v-L} and divided into two groups based on a threshold of 100 MHz.
%For a single burst, if we assume that its outflow region is approximately circular, we uniformly slice this region (see Appendix \ref{C}).
%We can obtain the image of frequency-luminosity in Figure \ref{v-L}, allowing for a more comprehensive analysis of the emission characteristics.
%In the upper panel, it appears that high-frequency slices exhibit an earlier outburst compared to medium and low-frequency slices.

For Figure \ref{v-L}, in the upper panel, the luminosity variation with frequency is shown for a slice, while the bottom panel represents the cumulative luminosity variation for one burst.
Meanwhile, in our simulation, the high-frequency bunches emit radiation earlier compared to the middle-frequency and low-frequency bunches.
Moreover, the overall luminosity distribution appears to exhibit a phenomenon of rapid brightening followed by a gradual dimming.
If we consider the possibility of a low-frequency cutoff and only focus on the middle and high-frequency component, the luminosity of a burst could potentially follow a power-law distribution.
Furthermore, if the frequency range corresponding to the rapid rise in luminosity falls within the observational range, our model may potentially explain the occurrence of bright spots during the burst.
Of course, it should be noted that the high, middle, and low frequencies are relative values rather than strict bands.
%Meanwhile, the mid-frequency bunches display the highest luminosity.
%Furthermore, we notice that the burst may be narrow-band, given the low number of high-frequency bunches and the abundance of low-frequency bunches with fainter luminosity and greater susceptibility to absorption.
Therefore, these findings suggest that, for an individual burst, there may exist phenomena of frequency downward drifting and bright spots.
Coincidentally, these features bear some resemblance to the observed phenomenon of FRB 20190520B.

\begin{figure*}[htbp]
\plotone{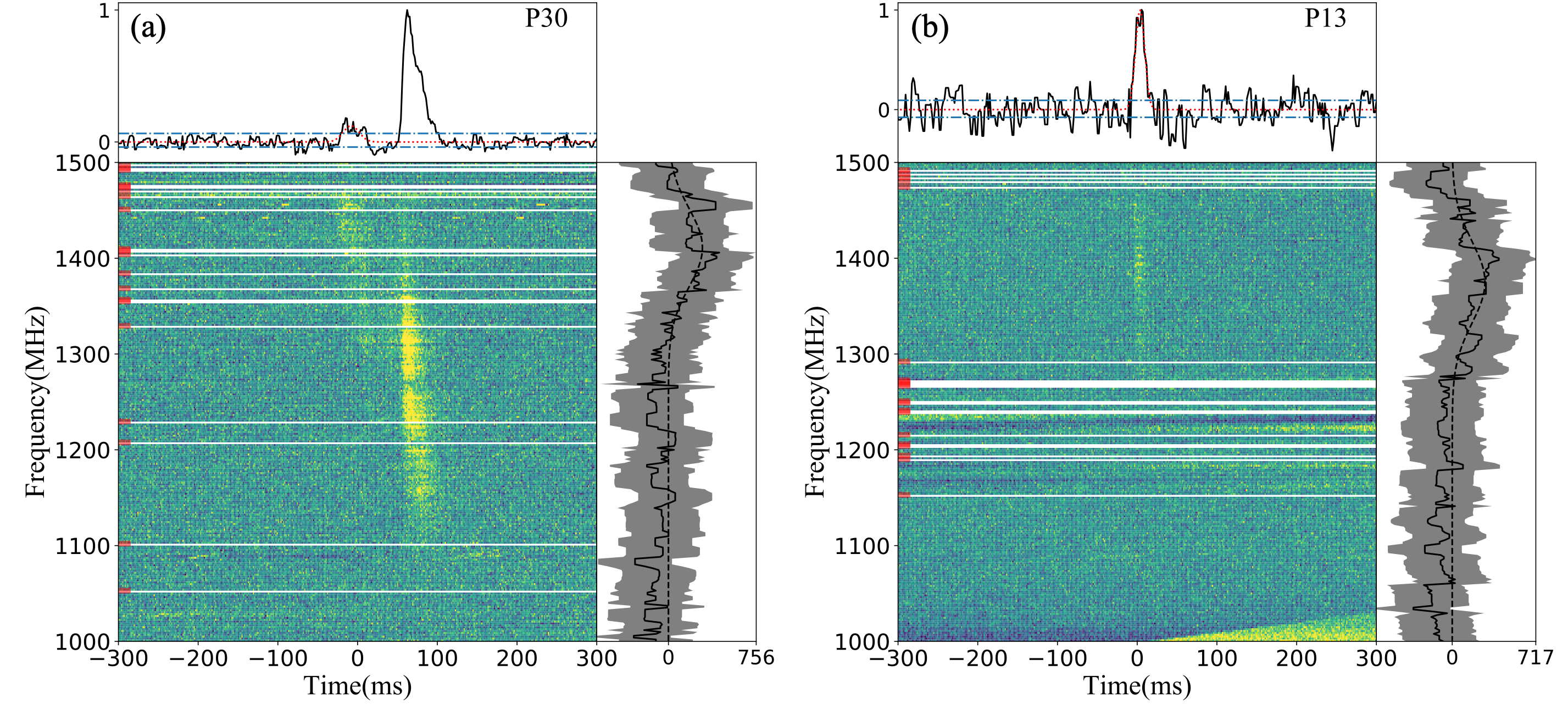}
\caption{Dynamic spectra of two bursts from FRB 20190520B.
The upper plots mean the pulse intensity varying with time. 
The bottom-left plots represent the dynamic spectrum after de-dispersion. 
The bottom-right plots show the intensity varying with frequency.
Red markers mean frequency channels of the radio frequency interference (RFI), and white horizontal lines indicate the areas covered by the RFI.
Sub-figure (a) shows the frequency downward drifting of the sub-burst; sub-figure (b) shows the bright spot that appears in the mid-frequency. 
\label{sd1}}
\end{figure*}

\subsection{Application for FRB 20190520B}\label{3.3}
FRB 20190520B, a compelling source discovered by FAST \citep{Niu22}, presents characteristics of active bursts, association with a PRS, and an $\rm DM_{IGM}$ that markedly deviates from the Macquart relation \citep{Macquart20}.
These phenomena earmark this source as a distinctive and representative FRB, thus analyzing this source from a theoretical perspective is important. 
From an energy burst perspective, FRB 20190520B manifests a log-normal distribution with a mean isotropic burst energy of around $\sim10^{38}\,erg$.
Assuming a duration of approximately $1\,ms$, the mean isotropic luminosity is around $10^{41}\,erg\,s^{-1}$, which falls within the given value of Eq.\ref{liso}.
Considering the higher-luminosity bursts within the dataset, which approach $10^{42}\,erg\,s^{-1}$, these also conform to the value indicated in Eq.\ref{liso}.
Concurrently, when considering each slice in Figure \ref{v-L} as having distinct luminosities, the cumulative luminosity approximates $\sim 1.3\times10^{43}\,erg\,s^{-1}$.
Limiting consideration to slices with a frequency above 100 MHz, the luminosity sum is $\sim 8.2\times10^{42}\,erg\,s^{-1}$, sufficient for an FRB 20190520B burst.
Even restricting the calculation to the FAST observation band (1-1.5GHz), the cumulative luminosities remain substantial at $\sim 7.0\times10^{41}\,erg\,s^{-1}$, thereby still meeting the observational requirements of FRB 20190520B.
Consequently, from a luminosity perspective, our model adequately satisfies observational demands.

Indeed, only focusing on luminosity or energy analysis does not entirely illustrate the strength of our model, since these aspects can be managed by adjusting model parameters. 
What truly showcases the value of our model is its may interpret the complex observational phenomena of FRB 20190520B, such as the frequency drifting and the localization of bright spots.
In regard to frequency drifting, we have noted that a downward-drifting pattern is frequently observed in FRBs \citep{Pleunis21, Zhou22}. 
\cite{Wang19} suggested that for a constant $\gamma$, this downward drifting feature may manifest in FRBs. 
Our model extends this notion by incorporating a variable $\gamma$, which can also give rise to a downward frequency drift phenomenon.
In the context of FRB 20190520B, we investigated 75 reported bursts and identified 12 pairs of sub-bursts with conspicuous downward drifting, akin to the patterns shown in sub-figure (a) in Figure \ref{sd1}. 
%while the remaining bursts either showed unclear drifting or were difficult to discern due to a low signal-to-noise ratio or exhibited upward drift.
If we postulate that different sub-bursts from one pair emanate from different components of the same burst, our model could offer a plausible explanation for this occurrence. 
As illustrated in Figure \ref{v-L}, high-frequency components are emitted earlier than their low-frequency counterparts. 
Further discussion is shown in Section \ref{4.3}.
%However, our model currently cannot well explain the up-drifting bursts. 
%Nonetheless, we also noticed that these bursts often have lower signal-to-noise ratios, implying that the up-drifting phenomenon may be related to complex surrounding environments.

Lastly, we have observed that for some bursts, the luminosities are not uniform in the time-frequency plot but feature bright spots, as exemplified in sub-figure (b) in Figure \ref{sd1}.
This phenomenon could be ascribed to scintillation, which results in conspicuous bright-dim stripe patterns in the dynamic spectrum \citep{Main22}.
However, if bright spots, instead of stripes, are observed, they might be associated with the FRB mechanism itself, a common occurrence in FRBs.
For instance, in the case of FRB 20190520B, roughly one-third of the bright spots manifest in the middle of the burst channel (mid-frequency), as opposed to the top (high-frequency) or bottom (low-frequency). 
In our simulation, as illustrated in Figure \ref{v-L}, we find that the luminosities around 500MHz outshine those in other frequency bands.
%As shown in Figure \ref{v-L}, the number of points or the luminosity is lower in the high-frequency and low-frequency regions. 
%In contrast, the number of points and the luminosity is higher in the mid-frequency region. 
Consequently, this could manifest in observational as bright spots occurring near mid-frequencies.
Meanwhile, for mid-frequencies around 500MHz, the CHIME telescope may offer a more suitable platform for validating this simulation.
Indeed, some bursts documented in CHIME/FRB Catalog 1 have exhibited a brightening phenomenon at mid-frequencies \citep{Pleunis21}.
It's worth noting that this simulation does not precisely match the current observations of FRB 20190520B.
For example, the mid-frequency range with the highest luminosity may differ.
Nevertheless, when examining the region of the FAST frequency in Figure \ref{v-L}, we can also discern luminosity fluctuations.
Thus, the luminosity variation in our model may provide a clue for explaining the the phenomenon of bright spots.

\section{Discussion} \label{4}
In this section, we discuss several related implications of our model, including model assumptions and the initial considerations for bunch formation.
Additionally, possible observational concerns are also addressed, which encompass frequency downward drifting, bright spots, narrow-band signals, and potential applicability to other FRBs.
\begin{figure*}[htbp]
\plotone{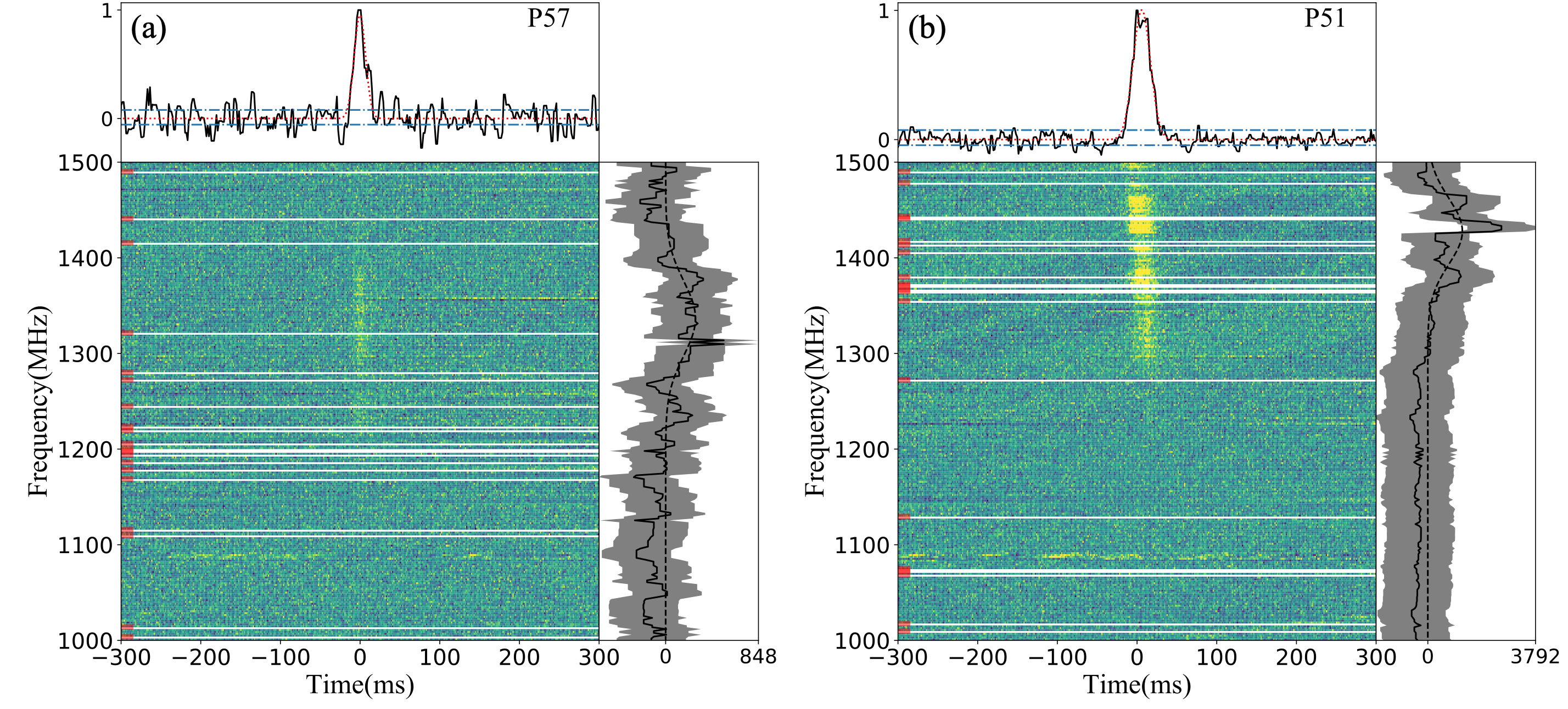}
\caption{Dynamic spectra of two bursts from FRB 20190520B.
Sub-figure (a) represents narrow-band signal; sub-figure (b) means the burst locates at the edge of telescope bandwidth.
Other annotations are consistent with those in Figure \ref{sd1}.
\label{sd2}}
\end{figure*}

\subsection{Propagation of FRBs}\label{4.1}
A fundamental assumption underlying our model is that the radiation emit from the inner magnetosphere. 
This assumption gives rise to a simple but significant question: can the radio signals effectively propagate through the surrounding medium?
If we make an assumption that the medium has a density of the Goldreich-Julian density \citep{Goldreich69} that 
\begin{equation}
    n_{GJ} \sim 10^{-2} B_s \Omega \sim 10^{16}B_{s,15}\Omega_{3}\,cm^{-3},
\end{equation}
where $\Omega$ is the rotation angular frequency of NS and we take $\Omega_{3}\sim 1000\,rad\,s^{-1}$, then the plasma frequency is
\begin{equation}
    \nu_{p,GJ} = \left(\frac{e^2 n_{GJ}}{\pi m_e}\right)^{1/2}\sim 900\,GHz.
\end{equation}
Moreover, presuming that the density of the surrounding medium is approximately $n_b\sim10^{17}\,cm^{-3}$, the corresponding $\nu_{p,b}\sim2.8\times10^{3}\,GHz$ is found. 
Notably, both $\nu_{p,GJ}$ and $\nu_{p,b}$ are significantly greater than the propagation frequency of FRBs ($\sim1\,GHz$).
Hence, the outflow region must either be located within the region of open magnetic field lines, rather than the co-rotating region, or a triggering mechanism must be in place to clear the surrounding medium and enable the propagation of FRB signals.
Additionally, several theories propose that as long as the condition $\nu_{FRB}>\nu_{p}^2/\nu_{cyc}$ is met, radio waves can penetrate high-density medium \citep{Arons86, Kumar17}, where $\nu_{cyc}\equiv eB_s/(2\pi m_ec)\sim2.8\times10^{12}B_{s,15}\,GHz$ is the electron cyclotron frequency.
From this viewpoint, even though $\nu_{FRB} \ll \nu_{p,GJ}$, GHz waves can still propagate that close to the speed of light.
Besides, it is worth noting that the radiation pressure may exceed the plasma pressure in the magnetosphere \citep{Wang22}, which could allow FRBs to break out of the surrounding medium.
Thus, based on the mentioned three perspectives, the propagation problem of FRB can be roughly addressed.

\subsection{Formation of bunches}\label{4.2}
Revisiting our initial question that can bunches be formed?
According to our model, bunches may dynamically form as a result of the electron deceleration mechanism, and the compression effect could result in the necessary particle number within the magnetic tube.
Whether such high-density bunches can form still needs further discussion, as many models tend to avoid addressing this issue.
To begin with, we can provide a rough phenomenological analysis from the perspectives of electrostatic potential energy and kinetic energy.
The electrostatic potential energy ($E_p$) can be viewed as the potential energy of all electrons in the bunch with respect to the electrons entering the bunch, which is written as 
\begin{equation}
    E_p = 3ke^2N_e/5r,
\end{equation}
where $k$ is Coulomb constant, and $r$ represents the distance between the bunch and the entering electron, which is similar to $l_b$. 
On the other hand, the particle kinetic energy ($E_k$) refers to the kinetic energy of the relativistic electron at the onset of ICS, which can be expressed as
\begin{equation}
    E_k = E_{etot} - E_0 = \left(m_e^2c^4+p_e^2c^2 \right)^{1/2} - m_ec^2,
\end{equation}
where $E_{etot}$ is total energy of a relativistic electron, $E_0=m_ec^2$ is the rest energy of the electron, and $p_e = \gamma m_ec$ is the momentum of the electron.
If the kinetic energy is greater than the potential energy, it can be approximated that the electrons have the ability to enter the bunch, thereby facilitating the compression process as we described.

\begin{figure}[htbp]
\plotone{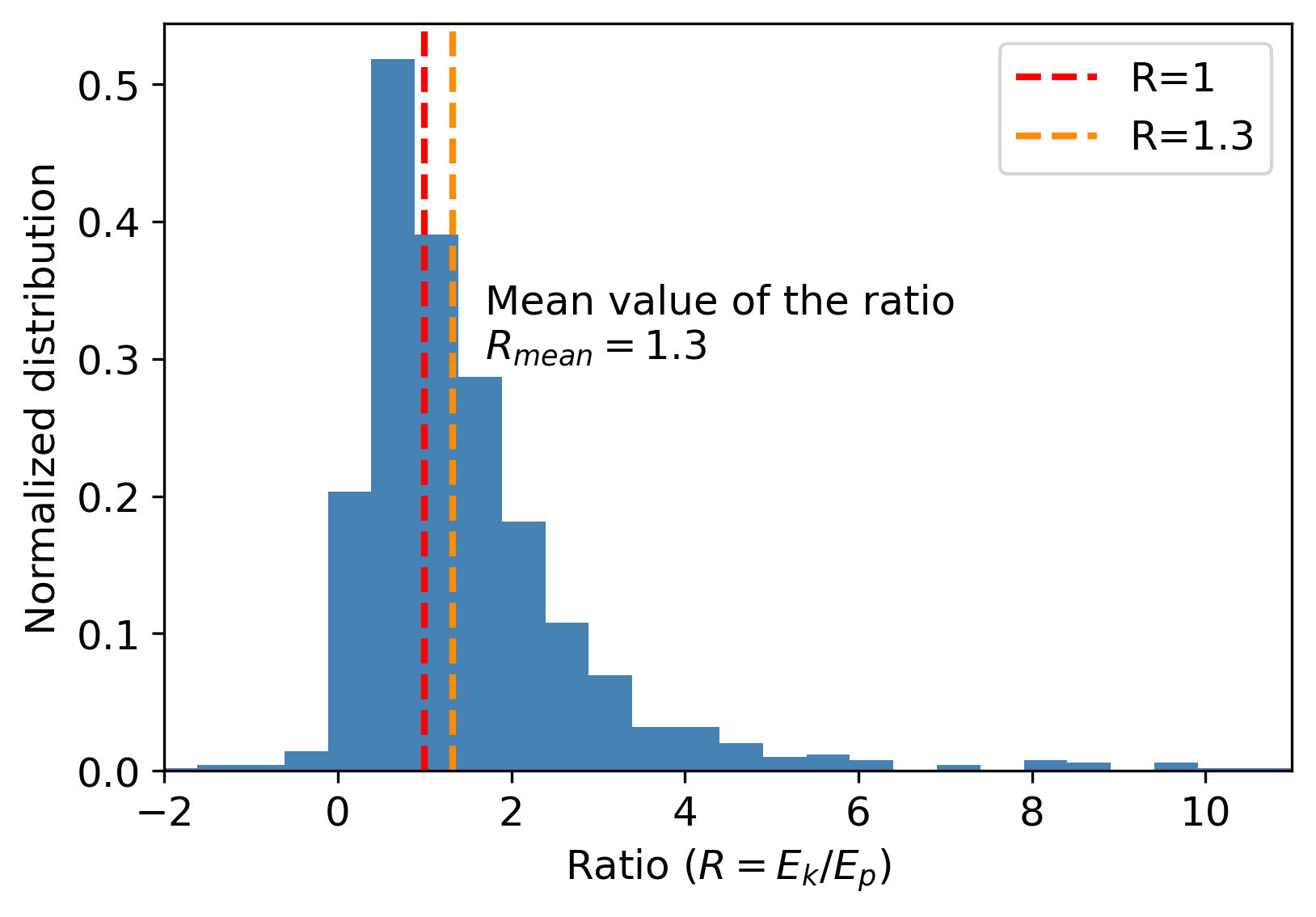}
\caption{The histogram of the ratio ($R = E_k/E_p$) normalized distribution based on Eq.\ref{ratio}.
The red and orange dashed lines represent the value of $R=1$ and $R=1.3$, respectively. 
\label{r-gau}}
\end{figure}

To estimate the balance between the two energies, statistical methods can be employed. 
Assuming that the parameters $N_e$, $r$, and $\gamma$ conform to a Gaussian distribution, a ratio ($R$) can be introduced to assess this matter as
\begin{equation}
    R = E_k(\mu_{\gamma}, \sigma_{\gamma}) / E_p(\mu_n,\mu_r, \sigma_n, \sigma_r),
\label{ratio}
\end{equation}
where $\mu$ and $\sigma$ are mean and variance in the Gaussian distribution.
For this, we can specify values for which $\mu_{\gamma} = 1000$, $\sigma_{\gamma} = 200$, $\mu_n = 10^{22}$, $\mu_r = 20\,cm$, $\sigma_n = 5\times10^{21}$, and $\sigma_r = 10\,cm$.  
We perform random sampling 1000 times for calculation, and the distribution is shown in Figure \ref{r-gau}.
When $R>1$, the kinetic energy surpasses the potential energy, implying that bunches can potentially form.
Meanwhile, as $R_{mean} = 1.3 >1$, the formation of bunches is feasible in the majority of cases.
To further mitigate the potential errors caused by different samplings each time, we cycle the aforementioned process 100 times, obtaining the mean value of the ratio for each times.
Based on these computed values, the harmonic mean value is calculated to be $R_{har} = 1.5$, closeing with the value depicted in Figure \ref{r-gau}.
Hence, from a statistical viewpoint, the formation of bunches appears plausible.

Furthermore, the magnetic pressure generated by the strong magnetic fields surrounding the bunches also plays a crucial role in confining these bunches.
From this perspective, we can estimate the relationship between the electrostatic potential ($E_p$) and the work done by the magnetic field ($W_B$).
A criterion for when bunches may disperse can be determined as follows:
\begin{equation}
    \Delta E_p > \Delta W_B,
\label{wb}
\end{equation}
where $\Delta E_p = 3ke^2/(5r^2)\,dr$ and $\Delta W_B = 4\pi P_Br^2\,dr$ represents the changes in electrostatic potential and the work done by the magnetic field, respectively. 
Assuming that the bunches are located at an altitude of $\sim 10^7\,cm$, when the surface magnetic field is $B_{s,15} = 10^{15}\,G$ and the neutron star (NS) radius is $R_{NS}\sim 10^6\,cm$, the magnetic field within the bunches is $B_{12} = 10^{12}\,G$, so the magnetic pressure $P_B \sim B_{12}^2/(8\pi)$.
When we substitute the parameters mentioned earlier, we find that the criterion outlined in Eq.\ref{wb} is not satisfied.
Therefore, the magnetic field should be sufficient to confine the bunches at this altitude.

According on the above discussions, we can summarize a few key conclusions.
Firstly, an energy-based analysis of the formation of bunches under our mechanism provides an intuitive understanding of the process.
Secondly, by examining the $E_k$, the $E_p$, and the $W_B$ under given parameters, we find that the formation of bunches is plausible.
Lastly, when we apply statistical methods to these considerations, the results support the likelihood of bunch formation within the defined parameter space.
Additionally, our model does not require the bunches to maintain a certain size after the FRB emission. 
%while flowing along the magnetic field. 
%In other words, there is a high probability that the scale of bunches will no longer be satisfied for coherent radiation after the FRB emission.
Simultaneously, given the exceedingly short radiation duration of each bunch, there is no requirement for them to sustain their structural integrity over an extended period \citep{Zhang22}.
%Therefore, such a dynamic model can avoid these issues to some extent.
However, it should be noted that the characteristics of bunches in the variable magnetosphere are quite complex, which needs further investigation, like considering electromagnetic fields, magneto-hydrodynamic waves, and particle-in-cell simulation, to make a complete understanding of this process.
Thus, our discussion represents merely an initial exploratory effort concerning this matter.
%Furthermore, due to the transient nature of coherent bunches in our model, the region of coherent radiation is limited to a relatively small area rather than throughout the entire outflow process.
%Moreover, active FRB bursts may require frequent energy triggers and dense soft photons, which could be indicative of a young celestial body such as a young magnetar.
%Thus, based on this model, the event rate of magnetars producing FRBs should decrease, which may help to explain the observational discrepancy \citep{Lin20} between the much higher event rate of magnetar flares and the rate of FRBs.

\subsection{Concerns of the application}\label{4.3}
From the standpoint of frequency downward drifting, although we have addressed it in Section \ref{3.3}, it needs additional consideration.
In our model, the prerequisites for a downward frequency drift are less stringent, not necessitating sub-bursts, and a single burst may also engender it.
This implies that within what is perceived as a single burst, multiple structures might exist, although they are challenging to distinguish.
It should, however, be noted that frequency drifting in a single burst might also be attributed to fluctuating dispersion measures.
Especially in the case of FRB 20190520B, with its complex environmental surroundings, further observational and analysis is required.

The ability of this model to explain other bursts with varying patterns, such as bright spots located in the low or high-frequency bands, is something to consider. 
We hypothesize that this could be connected to the structure of the outflow region. 
If we partition the region into several equal parts along the line of sight, their horizontal scales would be the same, but due to the shape of the region, their vertical scales would differ.
Consequently, the area of each slice would also vary, leading to a difference in the number of magnetic tubes contained in each slice, and this would influence the variation in the number of bunches that can be generated by the outflow particles.
In other words, if the number of bunches in a slice is relatively large, the luminosity will be higher than in other slices.
For a standard circle, assuming a line of sight from right to left as in Figure \ref{cycle}, our previous analysis suggests that the bright spot should be around the mid-frequency band.
However, if the region is a triangle with its apex pointing to the left, the spot should appear in the high-frequency band.
Conversely, if the apex is pointing to the right, the spot should be present in the low-frequency band.
Hence, based on this concept, the model could potentially account for different burst patterns and even structures with multiple peaks.
%which will be further refined in future work.

The issue of narrow-band signals is indeed a common challenge faced by present relativistic coherent radiation models.
For FRB 20190520B (as seen in sub-figure (a) of Figure \ref{sd2}), the majority of observed bursts have a frequency bandwidth that is less than half of the instrument bandwidth, i.e., less than approximately 250 MHz for FAST.
In our model, the maximum frequency bandwidth is around 3 GHz, which indicates a broadband model. 
However, a possible explanation is that if we assume that the bright part in Figure \ref{v-L} represents the observed bursts with high signal-to-noise ratios, it could be considered as a narrow-band signal.
% this part also conforms to the characteristic of frequency downward drifting.
Further clarification is also required regarding the definition of intrinsic narrow-band characteristics, which is necessary for accurately analyzing the radiation features of FRBs. 
Although certain studies on giant pulses of pulsars and FRBs have suggested that the narrow-band is $\Delta \nu / \nu_{tele} \sim 0.1$ \citep{Metzger19, Thulasiram21}, where $\nu_{tele}$ is the bandwidth of the telescope, this is still dependent on the instrument and not intrinsic to the source.
Even different telescope settings and parameters could affect the identification and classification of narrow-band signals \citep{Law17}.
Additionally, we find that some bursts of FRB 20190520B are located at the upper or lower edges of the telescope's bandwidth as seen in sub-figure (b) of Figure \ref{sd2}.
It remains uncertain whether there are signals beyond the frequency boundary in such cases, and further joint observations are needed to provide evidence. 
%which is important for the analysis of FRBs narrow-band properties.
If such situations do exist, it is important for us to gain a further understanding on the morphology and burst mechanism of FRBs.
%for those bursts with bright spots appearing at the edge of the frequency band, the true position of the spots should be in the mid-frequency of the burst rather than the high/low-frequency parts. This 

Finally, the question arises that can this model be extended to other FRBs? 
We believe that this is currently an open question due to the uncertain diversity in the origins and mechanisms of FRBs. 
This model has been applied to FRB 20190520B in order to illustrate its capabilities in accounting for observed phenomena, but it would be a premature claim to consider it as a comprehensive and universally applicable model for all FRBs.
However, some of the phenomena observed in FRB 20190520B have also been documented in other FRBs \citep{Pleunis21}, implying possible shared characteristics among them.
%such as frequency downward-drifting and bright spots.
For instance, the downward frequency drifting has been noted in many repeating FRBs, including FRB 20121102A \citep{Hessels19} and FRB 20201124A \citep{Zhou22}.
Moreover, for some FRBs, the flux of a single burst exhibits non-monotonic variation trends with fluctuations or bright spots \citep{Day20}, such as FRB 20180814A \citep{CHIME19} and FRB 20180916B \citep{Marcote20}.
Taken as a whole, these observations suggest that our model may provide some inspirations into explaining these commonly observed phenomena, particularly in relation to the compression effects of the bunch model.
%In the future work, we will further explore its application to other FRBs and multi-phenomena.

\section{Conclusion} \label{5}
In this work, we have expanded on the mechanism of FRBs, produced by bunches via coherent curvature radiation within a NS framework.
Our primary focus has been on the compression effect exerted by the ICS process on outflow particles.
We have provided a phenomenological discussion on the feasibility of bunch formation from an energy standpoint.
Upon further analysis of the corresponding observational effects, our model potentially explains some of the phenomena observed in FRB 20190520B. 
We wrap up our paper with a brief summary of key conclusions:
\begin{itemize}
    \item We have developed a model for the FRB mechanism rooted in coherent curvature radiation and magnetized NS.
    This model incorporates the process of bunch compression, thereby offering a novel approach towards understanding the formation of these bunches.
    \item We introduced the mechanisms of pair production and ICS preceding curvature radiation (FRB emission). 
    Our estimations suggest that ICS plays a significant role in compressing the bunches.
    Consequently, these bunches may dynamically form and emit curvature radiation along a curved magnetic field tube/family at specific positions.
    \item We discussed factors influencing the formation of bunches, encompassing the electrostatic potential energy within the bunch, the kinetic energy of relativistic electrons, and the work done by the magnetic field. 
    According to our analysis, the compressed bunch model fulfills the conditions required for the formation of bunches.
    \item We utilized this model to FRB 20190520B and successfully explained various observed phenomena, including some basic characteristics and the frequency downward drifting. 
    Furthermore, we provided insights into the position of bright spots in FRB 20190520B.
\end{itemize}

%\software{Python, numpy, scipy, matplotlib, math, astropy \citep{Astropy13}}
%% Also note that the akcnowlodgment environment does not support long amounts of text. If you have a lot of people and institutions to acknowledge, do not use this command. Instead, create a new \section{Acknowledgments}.

\begin{acknowledgments}
We thank Weiyang Wang, Xiaoping Zheng, Bifang Liu, Shiqi Zhou, Changqing Ye, and Yuhao Zhu for helpful discussions and suggestions from Shuangqiang Wang and Rui Luo.
This work is supported by the National Natural Science Foundation of China (Grant No. 11988101, 12203069, 12163001) and the National SKA Program of China/2022SKA0130100. 
This work is also supported by Office the leading Group for Cyberspace Affairs, CAS (No.CAS-WX2023PY-0102).
Finally, we thank the anonymous referee for the valueable comments and suggestions, which have significantly improved the quality of the paper.
%the International Partnership Program of Chinese Academy of Sciences grant No. 114A11KYSB20160008,
%the National Key R\&D Program of China No. 2016YFA0400702, and the Guizhou Provincial Science and Technology Foundation (Grant No. [2020]1Y019).
%No. U1938117, No. U1731238, No. 11703003, No. 11725313, and 12163001
\end{acknowledgments}





%\clearpage

\appendix
\section{Determination of model parameters }\label{A}
In this section, we will introduce the selection of parameters in the model, as well as the rationale for utilizing these values.
In summary, the parameters can be divided into three parts: neutron star (NS) parameters, particle outflow parameters, and FRB burst parameters.
Finally, we will provide a notation list as Table \ref{tab1}.
%that includes all the parameters relevant to the model, their definitions, and their first appearance.

Firstly, for the NS part, the parameters include the surface magnetic field $B_s$, the volume of magnetic energy release $V_m$, the energy conversion efficiency $\eta$, the initial Lorentz factor $\gamma_6$, and the surrounding temperature $T$.
Since an FRB-like signal has already been observed from Galactic magnetar SGR 1935+2154, it is reasonable to consider magnetars as the central powering objects. 
Moreover, the characteristic surface magnetic field of magnetars is typically in the range of $10^{14-15}\,G$, and considering that FRB bursts are expected to be relatively violent, selecting a value of $10^{15}\,G$ is justified.
Regarding the volume part, there are two parameters, the radius and depth of energy release that can be triggered by the mechanism, such as starquakes, magnetic reconnection, and explosions.
The selection of radius is arbitrary, and we assume it may be close to the polar cap area of the NS, although they are not necessarily the same, which is $R_e \sim 5\times10^4\,cm$.
The selection of depth is based on the balance between the magnetic pressure and the material pressure that $h_{c}=B_s^2/(8\pi\varrho_{nc}g_{ns})\sim4\times10^3\,cm$, where $\varrho_{nc}\sim10^{11}\,gcm^{-3}$ is the density of NS crust and $g_{ns}\sim 10^{14}\,cms^{-2}$ is the gravitational acceleration of NS.
Therefore, the corresponding volume is $V_m \sim 3\times 10^{13}\,cm^3$.
For the estimation of energy efficiency, radio efficiency cannot be directly used. 
Instead, the efficiency of conversion between matter and energy should be selected. 
Additionally, since the efficiency corresponding to starquakes and magnetic reconnection is still uncertain, we adopt the efficiency of a nuclear explosion $\eta\sim10^{-3}$.
For the initial Lorentz factor, we arbitrarily select a value of $\gamma_6\sim10^6$, which is close to the $\gamma$ of particles through the acceleration gap (although the meanings of the two are different). 
Some accretion NS systems can produce TeV photons \citep{North87}, which also corresponds to the value of $10^6$.
Next is the trigger time, which is selected based on the NS glitch time $t_{tri}\sim10\,s$ \citep{Ashton19}.
%It should be noted that our intention here is not to assert that the trigger mechanism is caused by glitches. 
%Instead, we aim to emphasize that the above parameter selection is ordinary and reasonable in high-energy astrophysics.
Finally, for NS temperature $T$, it is not the brightness temperature but the actual temperature of soft photons that was heated by the surface of NS and suffuse the inner magnetosphere, which we select a value of $T\sim10^6\,K$ \citep{Goldreich69}.

The second part concerns the outflow parameters, including the ratio of the number of particles from the trigger to the outflow region $\zeta$, the ratio of magnetic flux tube cross-section to outflow area $\kappa$, the Lorentz factor after cascading $\gamma_3$, and the compression time $\delta t$.  
The selection of $\zeta\sim10^{-3}$ is flexible because the particles generated by the triggering mechanism are isotropic, but the outflow region is outside the triggering region.
So the value is variable, but it should be small.
The parameter $\kappa \sim 10^{-8}$ is calculated by $S_{tube}/S_{flow}$, where $S_{tube} = 25 \pi \,cm^2$ is the cross-sectional area at the bottom of the magnetic tube and $S_{flow} = 25\times 10^8 \pi \,cm^2$ is the area of the outflow region.
Next is $\gamma_3\sim10^3$, which is related to the pair production and inverse Compton scattering (ICS) mechanisms, which will be discussed in Appendix B.
Finally, regarding the compression time $\delta t \sim10^{-4}\,s$, we referred to the integrated time of the ICS electron lifetime $t_{ics} \sim \gamma_{ics} m_e c^2 / P_{eic}$, where $\gamma_{ics}$ we take $\gamma_3 \rightarrow \gamma_{o,2}\sim 100$.

In the end, we will discuss the parameters related to the FRB region, containing the number of bunches $N_b$ and the curvature radius $\rho$.
Here, $N_b\sim3\times10^4$ is based on the average energy of FRB 20190520B, but this parameter may vary for other FRBs, so it is also flexible.
The selection of the $\rho$ needs to satisfy two conditions. 
To begin with, it should be larger than the radius of the NS ($\rho_{ns}\sim10^6\,cm$) because we need the open magnetic tubes in the outflow region. 
Then, it should be larger than the distance that particles move during the compression time ($\sim 3\times10^6\,cm$).
Moreover, considering that the compression does not start from the surface of the NS, selecting $\rho_7\sim 10^7\,cm$ is reasonable.

\begin{longtable}{*{3}{p{1cm}p{4.5cm}p{2cm}}}
\label{tab1}\\
\caption{Notation List.}
\\
\hline \hline \multicolumn{1}{l}{\textbf{Symbol}} & \multicolumn{1}{l}{\textbf{Definition}} & \multicolumn{1}{c}{\textbf{First Appear}} \\ \hline 
\endfirsthead

\multicolumn{3}{c}%
{\bfseries \tablename\ \thetable{}} {Continued} \\
\hline \multicolumn{1}{l}{\textbf{Symbol}} & \multicolumn{1}{l}{\textbf{Definition}} & \multicolumn{1}{c}{\textbf{First Appear}} \\ \hline 
\endhead

\hline \multicolumn{3}{r}{{To be Continued}} \\ \hline
\endfoot

\hline \hline
\endlastfoot
        $B$ & Magnetic field strength & Section \ref{2.2} \\
        $B_s$ & Surface magnetic field strength & Section \ref{2.2} \\
        $c$ & Speed of light & Section \ref{2.2} \\
        $\dot E_{loss}$ & Energy loss rate resulting in compression & Section \ref{2.2} \\
        $E_{0}$ & Rest energy of a electron & Section \ref{4.2} \\
        $E_{etot}$ & Total energy of a relativistic electron & Section \ref{4.2} \\
        $E_{k}$ & Kinetic energy of a relativistic electron & Section \ref{4.2} \\
        $E_{p}$ & Electrostatic potential energy & Section \ref{4.2} \\
        $E_{tot}$ & Total energy released & Section \ref{2.2} \\
        $e$ & Elementary charge & Section \ref{2.2} \\
        $g_{ns}$ & Gravitational acceleration of NS & Appendix \ref{A} \\
        $h_{c}$ & Depth of energy release & Appendix \ref{A} \\
        $\hbar$ & Reduced Planck constant & Appendix \ref{B} \\
        $k$ & Coulomb constant & Section \ref{4.2} \\
        $L_{iso}$ & Isotropic luminosity & Section \ref{2.2} \\
        $L_{isob}$ & Isotropic luminosity of a bunch & Section \ref{3.1} \\
        $l_b$ & Length of a bunch & Section \ref{2.2} \\
        $l_c$ & Length that can be compressed to $l_b$ for the particle stream  & Section \ref{2.2} \\
        $l_{str}$ & Length of the particle stream & Section \ref{2.2} \\
        $m_e$ & Electron/positron mass & Section \ref{2.2} \\
        $N_{cas}$ & Cascade number & Section \ref{2.2} \\
        $N_{e}$ & Particle number after cascade & Section \ref{2.2} \\
        $N_{tc}$ & Total outflow particle number & Section \ref{2.2} \\
        $N_{tot}$ & Initial particle number & Section \ref{2.2} \\
        $n_b$ & Number density in a bunch & Section \ref{2.2} \\
        $n_{GJ}$ & Goldreich-Julian (GJ) density & Section \ref{4.1} \\
        $P_{B}$ & Pressure of magnetic field & Section \ref{4.2} \\
        $P_{bic}$ & ICS power of a bunch & Section \ref{2.2} \\
        $P_{cr}$ & Power of curvature radiation & Section \ref{2.2} \\
        $P_{eic}$ & ICS power of a single electron & Section \ref{2.2} \\
        $p_{e}$ & Momentum of a electron & Section \ref{4.2} \\
        $R$ & Ratio between $E_k$ and $E_p$ & Section \ref{4.2} \\
        $R_{e}$ & Radius of energy release & Appendix \ref{A} \\
        $R_{har}$ & Harmonic mean value of Ratio & Section \ref{4.2} \\
        $R_{mean}$ & Mean value of ratio & Section \ref{4.2} \\
        $R_{NS}$ & NS radius & Section \ref{4.2} \\
        $r$ & Distance of electrostatic potential energy & Section \ref{4.2} \\
        $r_{0}$ & Classical radius of the electron & Appendix \ref{B} \\
        $S_{flow}$ & Area of the outflow region & Section \ref{2.2} \\
        $S_{tube}$ & Cross-sectional area of the magnetic tube & Section \ref{2.2} \\
        $T$ & NS temperature & Section \ref{2.2} \\
        $t_{ics}$ & ICS electron lifetime & Appendix \ref{A} \\
        $t_{tri}$ & Trigger time & Appendix \ref{A} \\
        $U_{ph}$ & Internal energy of the photon & Section \ref{2.2} \\       
        $V_m$ & Volume of the magnetic energy release & Section \ref{2.2} \\
        $v$ & Velocity of the outflow particle & Section \ref{2.2} \\
        $\alpha_i$ & Interaction angle between the incident photon and particle & Appendix \ref{B} \\
        $\beta$ & Speed ratio between the particle and light & Section \ref{2.2} \\
        $\gamma$ & Lorentz factor & Section \ref{2.2} \\
        $\gamma_o$ & Lorentz factor of burst & Section \ref{2.2} \\
        $\gamma_3$ & Lorentz factor after Cascade & Section \ref{2.2} \\
        $\gamma_6$ & Initial Lorentz factor & Section \ref{2.2} \\
        $\Delta E_p$ & Changes of electrostatic potential & Section \ref{4.2} \\
        $\Delta W_B$ & Changes of work done by magnetic field & Section \ref{4.2} \\
        $\Delta \nu$ & Radiation frequency bandwidth & Section \ref{4.3} \\
        $\Delta \nu_{max}$ & Maximum radiation frequency bandwidth for one burst & Section \ref{2.2} \\
        $\delta h$ & Height difference of various bunches & Section \ref{2.2} \\
        $\delta l$ & Compression length & Section \ref{2.2} \\
        $\delta t$ & Compression time & Section \ref{2.2} \\
        $\epsilon$ & Photon energy & Appendix \ref{B} \\
        $\zeta$ & Ratio of particle number & Section \ref{2.2} \\
        $\eta$ & Ratio of material-energy  & Section \ref{2.2} \\
        $\theta_i$ & Interaction angle between the scattered photon and incident photon & Appendix \ref{B} \\
        $\kappa$ & Ratio of area & Section \ref{2.2} \\
        $\lambda$ & Wavelength & Section \ref{2.2} \\
        $\mu_{\gamma, n, r}$ & Mean values of Gaussian distributions & Section \ref{4.2} \\
        $\nu_c$ & Characteristic frequency of curvature radiation & Section \ref{2.2} \\
        $\nu_{cyc}$ & Electron cyclotron frequency & Section \ref{4.1} \\
        $\nu_{FRB}$ & FRB frequency & Section \ref{4.1} \\
        $\nu_i$ & Initial frequency & Appendix \ref{B} \\
        $\nu_{p}$ & Plasma frequency  & Section \ref{4.1} \\
        $\nu_{p,b}$ & Plasma frequency with $n_b$ & Section \ref{4.1} \\
        $\nu_{p,GJ}$ & Plasma frequency with $n_{GJ}$ & Section \ref{4.1} \\
        $\nu_{tele}$ & Frequency bandwidth of telescope & Section \ref{4.3} \\
        $\xi$ & Ratio of volume & Section \ref{2.2} \\
        $\rho_{ns}$ & NS radius  & Appendix \ref{A} \\
        $\varrho_{nc}$ & Mass density of NS crust & Appendix \ref{A} \\
        $\sigma$ & Scattering cross-section & Appendix \ref{B} \\
        $\sigma_{\gamma, n, r}$ & Values Variance of Gaussian distributions & Section \ref{4.2} \\
        $\sigma_{sb}$ & Stefan-Boltzmann constant & Section \ref{2.2} \\
        $\sigma_T$ & Thomson scattering cross-section & Section \ref{2.2} \\
        $\Omega$ & Rotation angular frequency of NS & Section \ref{4.1} \\ 
        $\omega$ & Solid angle & Appendix \ref{B} \\

        \multicolumn{3}{l}{Subscript $i$ and $o$ represent the initial and end states of} \\
        \multicolumn{3}{l}{the physical parameters, respectively.}   \\

\end{longtable}

\section{Conditions of ICS}\label{B}
In our model, we require two effective energy dissipation processes to diminish the $\gamma$ of the electrons prior to the curvature radiation, namely, pair production and ICS. 
Pair production dissipates energy via cascades, whereas ICS does so by facilitating interactions between high-energy electrons and soft photons. 
Importantly, as $\gamma$ decreases, the mean free path of photons significantly increases \citep{Timokhin19}, leading to a substantial reduction in the efficiency of photon cascades during the pair production. 
Consequently, when ICS is dominant, such as when $\gamma=1000$, pair production can be approximately disregarded.

Next, we will analyze the conditions of ICS.
Considering a set of relativistic electrons in a thermal equilibrium photon field with a temperature of $T\sim10^6\,K$, the energy loss rate due to ICS of the electrons equals this radiation power, which is Eq. \ref{peic}.
%The ICS power is related to the energy of the soft photon, and the Thomson scattering cross-section ($\sigma_T$) can only be applied if $\gamma h \nu \ll m_e c^2$, where $\gamma h \nu$ is the photon energy in electron frame.
%which refers to the situation that in the reference of electrons frame, the photons under Lorentz transformation do not exceed the rest mass of the electrons,
This formula applies when the hot photos are not too hard, as $\gamma \hbar \nu \ll m_e c^2$. 
If not, such as $\gamma \sim \gamma_6$, the quantum effects should be considered, and the modified equation is \citep{Rybicki91}
 \begin{equation}
     P_{eic}^{\prime} = \frac{4}{3}\sigma_Tc\gamma^2\beta U_{ph}\left[1-\frac{63\gamma \bar{\epsilon^2}}{10m_e c^2 \Bar{\epsilon}}\right],
 \end{equation}
where the $\bar{\epsilon^2}$ and $\Bar{\epsilon}$ denotes the average of energy squared and energy, respectively. 
At extreme relativity situation, the scattering cross-section in unit solid angle should be described by Klein-Nishina formula:
\begin{equation}
     \frac{d\sigma}{d\omega} = \frac{r_0^2\epsilon_o^2}{2\epsilon_i^2}\left(\frac{\epsilon_i}{\epsilon_o} +\frac{\epsilon_o}{\epsilon_i} -sin^2\theta \right),
\end{equation}
where $r_0 \approx 2.8 \times 10^{-13}\,cm$ is the classical radius of an electron, the $\epsilon_o$ and $\epsilon_i$ denotes the energy of final state and initial state of photon, respectively, and $\theta$ is the angle between the scattered and incident photo. 
The scattering cross-section drops quickly at high frequency, and the energy change of each photon is small. 
Apparently, the ICS is not an effective energy-losing mechanism under this condition.
However, as a decrease of $\gamma$, the cross-section will increase to $\sigma_T$.
When we consider that soft photons come from the surface of a NS, then the applicable condition of ICS with $\sigma_T$ is
\begin{equation}
    \gamma \hbar \nu_i (1-cos\alpha_i)\ll m_ec^2.
\end{equation}
which gives the restriction relation between the initial energy and the initial incident angle ($\alpha_i$) of the photon. 
By considering the photons only coming from the back that $\pi/2 < \alpha_i <\pi$, and the energy of hot photons is about 100eV, the condition of $\gamma \ll 5\times10^3$ is required.
Meanwhile, in the relativistic electron frame, ICS efficiency is still low for soft photons coming from directly behind the electron, but the angle is relatively small $\sim 1/\gamma$, which can be ignored.
Thus, it is reasonable for us to choose $\gamma_3\sim10^3$ as the end of pair production and the beginning of ICS.

%\newpage
%\clearpage

\bibliography{model}{}
\bibliographystyle{aasjournal}


\end{CJK*}
\end{document}